\documentclass[11pt,a4paper]{article}
\usepackage{jheppub}
\usepackage{amssymb}
\usepackage{graphics}
\usepackage{epsfig}
\usepackage{slashed}

\begin{document}

\newcommand{\ppZHj}{$pp \to Z^0H^0j+X$ }
\newcommand{\ppWHj}{$pp \to W^{\pm}H^0j+X$ }
\newcommand{\ppWZHj}{$pp \to W^{\pm}(Z^0)H^0j+X$ }
\newcommand{\ppbarWHJ}{$PP/P\bar{P} \to W^{\pm}H^{0}+jet+X$}
\newcommand{\WH}{$W^{\pm}H^{0}+jet~$}
\newcommand{\ZH}{$Z^{0}H^{0}+jet~$}


\title{ Next-to-leading order QCD predictions for $Z^0 H^0 + {\rm jet}$ production at LHC }
\author{Su Ji-Juan,}
\author{Guo Lei,}
\author{Ma Wen-Gan,}
\author{Zhang Ren-You,}
\author{Ling Liu-Sheng,}
\author{Han Liang }
\affiliation{Department of Modern Physics, University of Science and
Technology of China (USTC), Hefei, Anhui 230026, P.R.China}
\emailAdd{sujijuan@mail.ustc.edu.cn}
\emailAdd{guolei@mail.ustc.edu.cn} \emailAdd{mawg@ustc.edu.cn}
\emailAdd{zhangry@ustc.edu.cn} \emailAdd{hanl@ustc.edu.cn}
\emailAdd{llsheng@mail.ustc.edu.cn} \emailAdd{hanl@ustc.edu.cn}

\abstract{ We calculate the complete next-to-leading order (NLO) QCD
corrections to the $Z^0H^0$ production in association with a jet at
the LHC. We study the impacts of the NLO QCD radiative corrections
to the integrated and differential cross sections and the dependence
of the cross section on the factorization/renormalization scale. We
present the transverse momentum distributions of the final $Z^0$-,
Higgs-boson and leading-jet. We find that the NLO QCD corrections
significantly modify the physical observables, and obviously reduce
the scale uncertainty of the LO cross section. The QCD K-factors can
be $1.183$ and $1.180$ at the $\sqrt{s}=14~TeV$ and $\sqrt{s}=7~TeV$
LHC respectively, when we adopt the inclusive event selection scheme
with $p_{T,j}^{cut}=50~GeV$, $m_H=120~GeV$ and
$\mu=\mu_r=\mu_f=\mu_0\equiv\frac{1}{2}(m_Z+m_H)$. Furthermore, we
make the comparison between the two scale choices, $\mu=\mu_0$ and
$\mu=\mu_1=\frac{1}{2}(E_{T}^{Z}+E_{T}^{H}+ \sum _{j}E_{T}^{jet})$,
and find the scale choice $\mu=\mu_1$ seems to be more appropriate
than the fixed scale $\mu=\mu_0$. }




\date{}
\maketitle
\flushbottom

\renewcommand{\theequation}{\arabic{section}.\arabic{equation}}
\renewcommand{\thesection}{\arabic{section}.}
\newcommand{\nb}{\nonumber}

\newcommand{\Dir}{\kern -6.4pt\Big{/}}
\newcommand{\Dirin}{\kern -10.4pt\Big{/}\kern 4.4pt}
\newcommand{\DDir}{\kern -7.6pt\Big{/}}
\newcommand{\DGir}{\kern -6.0pt\Big{/}}

\makeatletter      
\@addtoreset{equation}{section}
\makeatother       

\section{Introduction}
\par
The standard model (SM) is a non-Abelian gauge theory and most of
its predictions have been perfectly confirmed by the present precise
experimental data except the Higgs-boson, which explains the
mechanism of mass generation and is believed to be responsible for
the breaking of the electroweak symmetry \cite{1,AA1,AA2,AA3}.
Therefore, one of the primary tasks for the CERN Large Hadron
Collider (LHC) is to discover the Higgs-boson. Correspondingly,
major efforts have been concentrated on devising method for
Higgs-boson search. The four LEP collaborations have established the
lower bound of the SM Higgs mass as $114.4~GeV$ at the $95\%$
confidence level \cite{2}. The CDF and D0 experiments at the
Fermilab Tevatron have ruled out the SM Higgs boson with mass
between $156$ and $177~GeV$ at $95\%$ confidence level (CL)
\cite{mh-Fermilab}. The ATLAS and CMS experiments at the LHC have
excluded most of the $m_H$ ranges $146-466~GeV$ and $145-400~GeV$ in
latest reports \cite{mh-Atlas} and \cite{mh-CMS} respectively, at
$95\%$ CL. The present LHC data are not yet sensitive
to the region around $m_H \sim 120~GeV$ favored by the SM fit to
electroweak precision data \cite{mh-LHC}. Not long ago, the ATLAS
and CMS experiments at the LHC have provided the upper limit of the
SM Higgs mass as $130~GeV$ and $127~GeV$ at $95\%$ CL
respectively, and there are several Higgs like events around the
locations of $m_H \sim 126~GeV$(ATLAS) and $m_H \sim 124~GeV$ (CMS)
\cite{mh-Atlas-1}\cite{mh-CMS-1}. Further searching for Higgs boson and
studying the phenomenology concerning its properties are still the
important tasks for the present and upcoming high energy colliders.

\par
There are several Higgs search channels, such as: gluon fusion
($gg \to H^0$) induced mainly by a top-quark loop \cite{9,10},
weak boson fusion (WBF) \cite{11}, top-quark associated production
($t\bar{t}H^0$) \cite{12}, and weak boson associated production
($VH^0(V=W^{\pm},Z^0)$) \cite{13,14,15}. Among these channels, the
$V H^0~(V=W^{\pm}, Z^0)$ associated production processes are
promising channels at low Higgs mass region. Although the
production rates of the $VH^0~(V=W^{\pm},Z^0)$ associated
production processes are several times lower than that of the
gluon fusion channel, these association productions become
available for Higgs searching by using the techniques of jet
reconstruction, b-tagging and lepton identification
\cite{tag1a,tag1b,tag2,tag3}. Furthermore, the $VH^0$
$(V=W^{\pm},Z^0)$ associated production processes also provide
unique information on the couplings between Higgs and vector
gauge-bosons. Recently, the calculations of the QCD ${\cal
O}(\alpha_s)$ and electroweak ${\cal O}(\alpha_{ew})$ corrections
to the Higgs production processes $p\bar{p}/pp \to
W^{\pm}H^0/Z^0H^0+X$ at the Tevatron and LHC were presented in
Refs.\cite{QCDc1,QCDc2,QCDc3, QCDc4,QCDc5} and Ref.\cite{15},
respectively. The NNLO QCD corrections to the SM Higgs-boson
production processes associated with a vector boson at hadron
colliders have been calculated in Ref.\cite{13,nnlo1}. In
Ref.\cite{nnlo1} it concluded that the NNLO QCD corrections to the
$Z^0H^0/W^{\pm}H^0$ production processes at the Tevatron and the
LHC increase the cross sections by the order of $5-10\%$ , while
the electroweak ${\cal O}(\alpha_{ew})$ corrections have been
turned out to be negative and about $-5\%$ or $-10\%$ depending on
whether the weak couplings are derived from $G_{\mu}$ or
$\alpha_{ew}(m_Z^2)$, respectively \cite{15}. Therefore, after the
inclusion of these corrections, the remaining uncertainties for
the $Z^0H^0/W^{\pm}H^0$ production processes should be dominated
by factorization/renormalization scale dependence and parton
distribution functions.

\par
As we known, the experimental environment at hadronic collider is
extremely complicated. The $V H^0~(V=W^{\pm}, Z^0)$ associated
production signals at the LHC are normally accompanied by
multi-jets in final state. Therefore, it is very necessary to have
a good understanding of these multi-body final state processes.
That requires sufficiently precise predictions for the $V
H^0~(V=W^{\pm}, Z^0)$ associated production signals and their
backgrounds with multi-jets in final state which cannot entirely
be separated in experimental data. Actually, the inclusive $V
H^0~(V=W^{\pm}, Z^0)$ associated production signals include any
number of additional jets unless otherwise stated. In this sense
the $V H^0 + {\rm jet}~(V=W^{\pm}, Z^0)$ productions are part of
the inclusive $V H^0~(V=W^{\pm}, Z^0)$ productions, and
theoretically $V H^0 + {\rm jet}$ production at the
next-to-leading order (NLO) QCD is part of the $VH^0$ production
process at the QCD next-to-next-to-leading order (NNLO). The
process \ppWHj at hadron colliders including the complete NLO QCD
corrections, which are part of the NNLO QCD correction to the
$pp/p\bar{p} \to W^{\pm}H^0+X$ process, have been calculated in
Ref.\cite{me}.

\par
In this paper, we present the calculations for the process \ppZHj
at the LHC up to the NLO in the QCD. The paper is organized as
follows: In Sec.2, we describe the calculations of the tree-level
cross section and the QCD NLO corrections for the \ppZHj process
at the LHC. The numerical results and discussions are presented in
Sec.3. A short summary is given in Sec.4. Finally, we provide some
analytical expressions for the amplitudes of the LO and real
emission partonic processes in Appendix.

\vskip 5mm
\section{Calculations  }
\par

\par
{\bf  A. LO calculations }
\par
\begin{figure*}
\begin{center}
\includegraphics[scale=1.0]{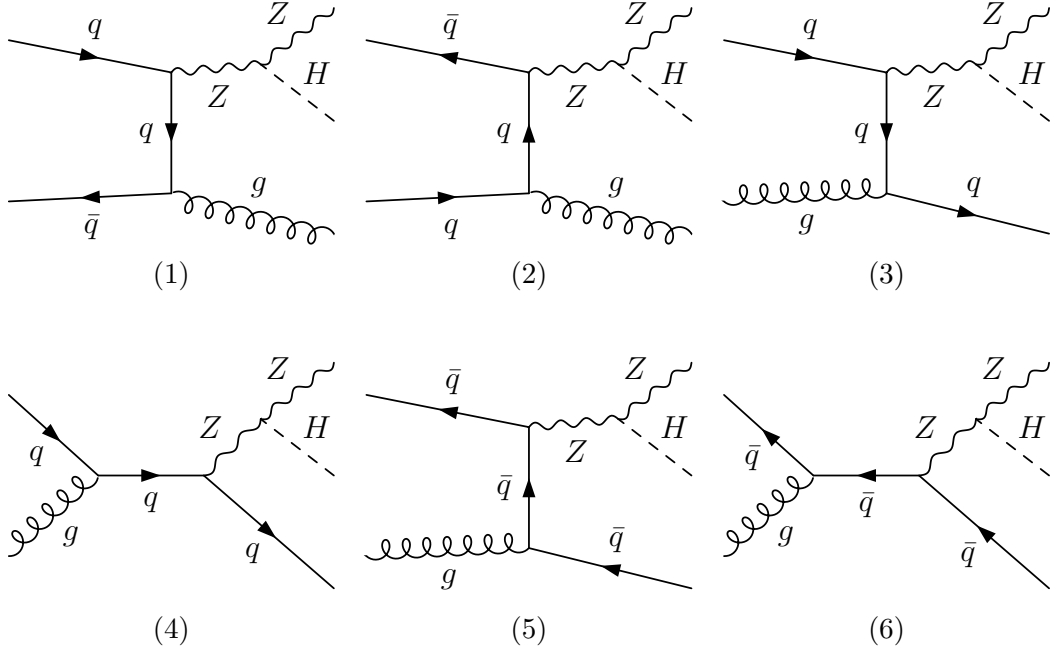}
\caption{\label{fig1} The generic LO Feynman diagrams for the
partonic processes $q\bar{q} \to Z^0H^0g$, $qg\to Z^0H^0 q$ and
$\bar{q}g\to Z^0H^0 \bar{q}$. (1) and (2) are the LO diagrams for
the partonic process $q\bar{q} \to Z^0H^0g$, (3) and (4) for
the $qg\to Z^0H^0q$, (5) and (6) for the $\bar{q}g\to Z^0H^0 \bar{q}$,
where $q=u,d,c,s,b$.  }
\end{center}
\end{figure*}

The LO cross section for the \ppZHj parent process involve following
partonic processes:
\begin{eqnarray}
\label{process-1} q~(p_{1})+\bar{q}~(p_{2})\to
Z^0~(p_{3})+H^0~(p_{4})+g~(p_{5}),  \\
\label{process-2}q~(p_{1})+g~(p_{2})\to Z^0~(p_{3})+H^0~(p_{4})+q~(p_{5}),   \\
\label{process-3}\bar{q}~(p_{1})+g~(p_{2})\to
Z^0~(p_{3})+H^0~(p_{4})+\bar{q}(p_{5}),
\end{eqnarray}
where $q=u,d,c,s,b~$ and $p_{i}~(i=1,...,5)$ represent the
four-momenta of the incoming partons and the outgoing $Z^0,~H^0$,
jet, respectively. There are 6 tree-level Feynman diagrams for the
partonic processes of the $Z^0H^0$+jet production,
(\ref{process-1})-(\ref{process-3}), shown in Fig.\ref{fig1}. There
Figs.\ref{fig1}(1-2) are the LO diagrams for the partonic process
$q\bar{q} \to Z^0H^0g$,  Figs.1(3-4) for $qg\to Z^0H^0 q$ and
Figs.1(5-6) for $\bar{q}g\to Z^0H^0 \bar{q}$. They can be grouped
into two different topologies. Figs.\ref{fig1}(1,2,3,5) belong to
u(t)-channel, while Figs.\ref{fig1}(4,6) to s-channel. Actually, the
partonic processes of the $Z^0H^0$+jet production at the LHC are
related to the amplitude of $0 \to Z^{0}H^0q \bar{q}g$ by crossing
symmetry at the LO.

\par
The LO cross section expressions for the partonic
processes $q\bar{q} \to Z^0H^0g$, $qg\to Z^0H^0 q$ and
$\bar{q}g\to Z^0H^0 \bar{q}$ have the forms respectively as:
\begin{eqnarray}\label{sigma_qqgg}
\hat{\sigma}^{q\bar{q}}_{LO}= \frac{1}{4}\frac{1}{9}\frac{(2 \pi
)^4}{2\hat{s}} \int \sum_{spin}^{color}|{\cal M}_{LO}^{q\bar{q}}|^2
d\Omega_{3}^{q\bar{q}},~~~ \hat{\sigma}^{qg(\bar{q}g)}_{LO}=
\frac{1}{4}\frac{1}{24}\frac{(2 \pi )^4}{2\hat{s}}\int
\sum_{spin}^{color} |{\cal M}_{LO}^{qg(\bar{q}g)}|^2
d\Omega_{3}^{qg(\bar{q}g)}~, \nb \\
\end{eqnarray}
where ${\cal M}_{LO}^{q\bar{q}}$, ${\cal M}_{LO}^{qg}$ and ${\cal
M}_{LO}^{\bar{q}g}$ are the amplitudes of the corresponding
tree-level diagrams for the partonic processes $q\bar{q} \to
Z^0H^0g$, $qg\to Z^0H^0 q$ and $\bar{q}g\to Z^0H^0 \bar{q}$
$(q=u,d,c,s,b)$ shown in Fig.\ref{fig1}. The factors $\frac{1}{4}$
and $\frac{1}{9}(\frac{1}{24})$ in Eqs.(\ref{sigma_qqgg}) are due to
spin- and color-averaging for the initial partons, respectively.
$\hat{s}$ is the partonic center-of-mass energy squared. The
summations in Eqs.(\ref{sigma_qqgg}) are taken over the spins and
colors of all the relevant particles in these partonic processes.
$d\Omega_{3}^{kl}$ ($kl=q\bar{q},qg,\bar{q}g$) in
Eqs.(\ref{sigma_qqgg}) is the three-body phase space element
expressed as
\begin{eqnarray}\label{PSelement}
d \Omega_3^{kl} =  \delta^4 (p_1+p_2-\sum_{i = 3}^5 p_i)
\prod_{j=3}^5 \frac{d^3 \vec{p}_j}{(2\pi)^3 2E_j}.
\end{eqnarray}

\par
The LO total cross section for the parent process $pp \to Z^0H^0 j+X$ can be expressed as
\begin{eqnarray}\label{sigma_PP}
\sigma_{LO}(AB(pp)\to Z^0H^0j+X) ~~~~~~~~~~~~~~~~~~~~~~~~~~~~~~~~~~~~~~~~~~~~~ \nonumber \\
~~~~~~~=\sum_{kl}^{} \int dx_A dx_B \left[ G_{k/A}(x_A,\mu_f)
G_{l/B}(x_B,\mu_f)\hat{\sigma}^{kl}_{LO}(x_{A}x_{B},\mu_f)+(A\leftrightarrow
B)\right].
\end{eqnarray}
Here $kl=q\bar{q},qg,\bar{q}g$, ($q=u,d,c,s,b$), $AB = pp$ and
$\mu_f$ is the factorization scale. $x_{A}(x_{B})$ describes the
probability to find a parton $k(l)$ in proton $A(B)$ defined as
\begin{eqnarray}\label{1}
x_{A}=\frac{p_{1}}{P_{A}},~~x_{B}=\frac{p_{2}}{P_{B}}.
\end{eqnarray}
Here $P_{A}$ and $P_{B}$ are the four-momenta of the corresponding
protons. $G_{k(l)/A(B)}$ are the LO parton distribution functions
(PDFs) for parton $k(l)$ in a proton.

\par
{\bf  B. NLO QCD corrections }
\par
Throughout our calculations, we take the 't Hooft-Feynman gauge
except when we verify the gauge invariance. In the NLO QCD
calculations we adopt the dimensional regularization (DR) method in
$D=4-2\epsilon$ dimensions to isolated the UV and IR singularities
and the modified minimal subtraction $(\overline{MS})$ scheme to
renormalize the colored fields. The one-loop diagrams are
essentially obtained from the tree-level diagrams of related
partonic processes (\ref{process-1})-(\ref{process-3}) and generated
by means of the FeynArts3.5 package \cite{fey}. The amplitudes are
further analytically simplified by the modified FormCalc programs
\cite{form}. The reduction of a tensor integral to the lower-rank
tensor and further to scalar integral is done with the help of the
LoopTools library \cite{form,Passarino} and the FF package
\cite{van}. There the dimensionally regularized 3- and 4-point
integrals with soft or collinear singularity have been added to this
library \cite{R.K,Dittmaier}. The output amplitudes are
numerically evaluated by using our developed Fortran subroutines for
calculating N-point integrals and extracting the remaining finite
$\epsilon \times \frac{1}{\epsilon}$ terms.

\par
The NLO QCD corrections to the parent process \ppZHj involve
following components:
\begin{itemize}
\item The virtual contribution to the partonic channels (\ref{process-1}),
(\ref{process-2}) and (\ref{process-3}) from the QCD one-loop and the corresponding
counterterm diagrams.
\item The contribution of the real gluon emission partonic processes.
\item The contribution of the real light-(anti)quark emission partonic processes.
\item The corresponding contribution of the PDF counterterms.
\end{itemize}

\par
The one-loop QCD contribution to the $pp \to Z^0H^0j+X$ process
contains both ultraviolet (UV) and soft/collinear infrared (IR)
singularities, but the total virtual correction contributed by both
the one-loop QCD and the corresponding counterterm diagrams is UV
finite after performing the renormalization procedure. Nevertheless,
it still contains soft/collinear IR singularity which can be
canceled by adding the contributions of the real emission partonic
processes and the corresponding PDF counterterms.

\par
The virtual corrections to the partonic processes $q\bar{q}\to
Z^0H^0g $, $qg \to Z^0H^0q$ and $\bar{q}g \to Z^0H^0\bar{q}$ can be
expressed as
\begin{eqnarray}
&& d\hat{\sigma}_V^{q\bar{q}}
    = \frac{1}{4}\frac{1}{9}\frac{(2 \pi)^4}{2\hat{s}}
    \sum_{spin}^{color}2 {\cal R}e
    \left[{\cal M}_{LO}^{q\bar{q}\dag}{\cal M}_V^{q\bar{q}}
    \right]d \Omega_{3}^{q\bar{q}},  \nonumber \\
 && d\hat{\sigma}_V^{qg(\bar{q}g)}
    = \frac{1}{4}\frac{1}{24}\frac{(2 \pi)^4}{2\hat{s}}
    \sum_{spin}^{color}2 {\cal R}e
    \left[
    {\cal M}_{LO}^{qg(\bar{q}g)\dag}{\cal M}_V^{qg(\bar{q}g)}
    \right] d \Omega_{3}^{qg(\bar{q}g)},
\end{eqnarray}
where ${\cal M}_{LO}^{q\bar{q}}$, ${\cal M}_{LO}^{qg}$ and ${\cal
M}_{LO}^{\bar{q}g}$ are the LO matrix elements of the partonic
processes $q\bar{q}\to Z^0H^0 g $, $qg \to Z^0H^0 q$ and $\bar{q}g
\to Z^0H^0 \bar{q}$ separately, and  ${\cal M}_{V}^{q\bar{q}}$,
${\cal M}_{V}^{qg}$ and ${\cal M}_{V}^{\bar{q}g}$ are the NLO QCD
virtual matrix elements of the corresponding partonic processes.

\par
The real gluon/light-(anti)quark emission partonic processes are
obtained from $0 \to Z^0H^0ggq\bar{q}$ and $0\to
Z^0H^0q\bar{q}q'\overline{q'}$ by all possible crossings of
(anti)quarks and gluons into the initial state. All the real
gluon/light-(anti)quark emission partonic processes are presented in
the following:
\begin{eqnarray} \label{eq-n}
&&  (1)~g(p_{1})~g(p_{2}) \to  Z^0(p_{3})~H^0(p_{4})~q(p_{5})~\bar{q}(p_{6}), \\
&&  (2)~q(p_{1})~\bar{q}(p_{2}) \to  Z^0(p_{3})~H^0(p_{4})~g(p_{5})~g(p_{6}), \\
&&  (3)~q(p_{1})~g(p_{2}) \to  Z^0(p_{3})~H^0(p_{4})~q(p_{5})~g(p_{6}), \\
&&  (4)~\bar{q}(p_{1})~g(p_{2}) \to  Z^0(p_{3}~H^0(p_{4})~\bar{q}(p_{5})~g(p_{6}), \\
&&  (5)~q(p_{1})~\bar{q}(p_{2}) \to  Z^0(p_{3})~H^0(p_{4})~q'(p_{5})~\bar{q'}(p_{6}), \\
&&  (6)~q(p_{1})~\bar{q'}(p_{2}) \to  Z^0(p_{3})~H^0(p_{4})~q(p_{5})~\bar{q'}(p_{6}), \\
&&  (7)~q(p_{1})~q'(p_{2}) \to  Z^0(p_{3})~H^0(p_{4})~q(p_{5})~q'(p_{6}), \\
&&  (8)~\bar{q}(p_{1})~\bar{q'}(p_{2}) \to
Z^0(p_{3})~H^0(p_{4})~\bar{q}(p_{5})~\bar{q'}(p_{6}),
\end{eqnarray}
where $q,q'=u,d,c,s,b$. There are totally 95 real emission partonic
channels, and these tree-level partonic processes contain both soft
and collinear IR singularities. After the summation of the
renormalized virtual corrections with all the real partonic emission
corrections, the result is soft IR-safe, but still contains remained
collinear divergence. It will be totally IR-safe when we include the
contributions from the collinear counterterms of the PDFs.

\par
\begin{figure*}
\begin{center}
\includegraphics*{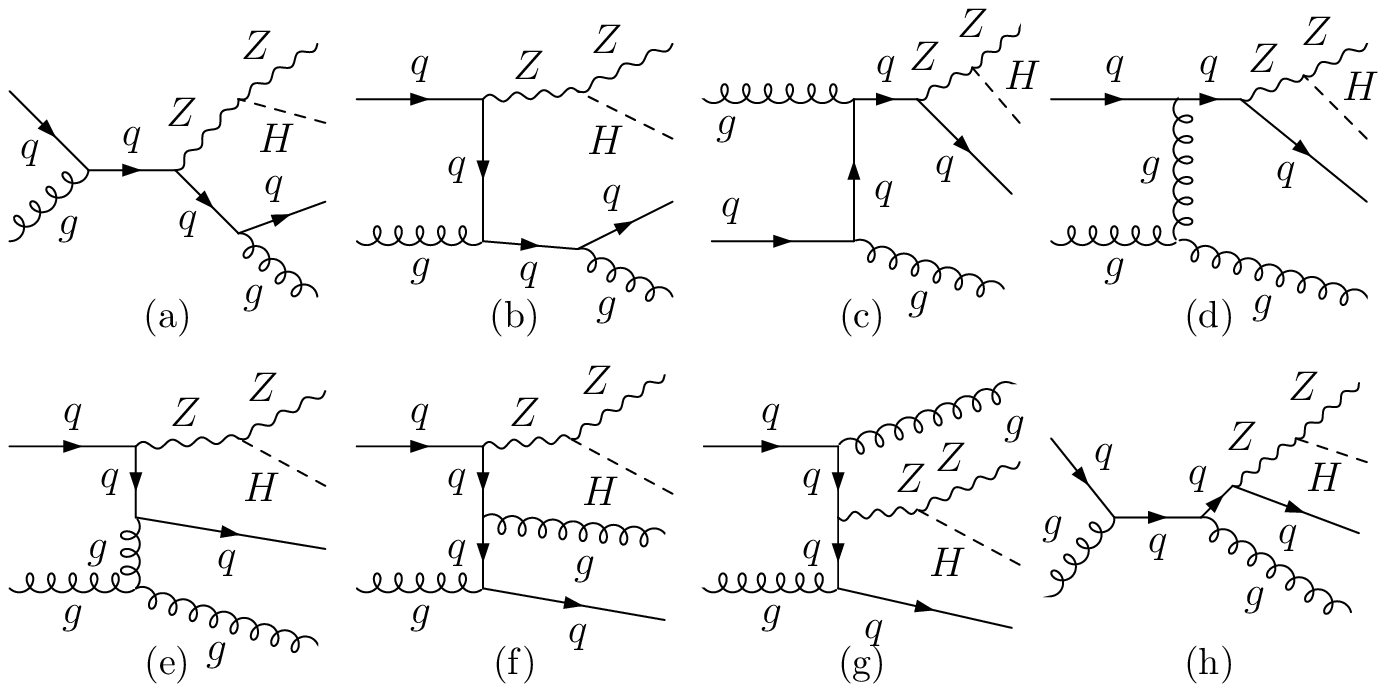}
\caption{\label{fig2} The tree-level Feynman diagrams for the real
emission partonic process $qg \to Z^0H^0qg$. }
\end{center}
\end{figure*}

As a demonstration we show in Figs.\ref{fig2}(a-h) the tree-level
Feynman diagrams for the real emission partonic process $qg \to
Z^0H^0qg$. We adopt the two cutoff phase space slicing (TCPSS)
method \cite{tcp} to isolate the IR singularities by introducing two
cutoff parameters $\delta_{s}$ and $\delta_{c}$. An arbitrary small
$\delta_{s}$ separates the four-body final state phase space into
two regions: the soft region ($E_{6}\leq
\delta_{s}\sqrt{\hat{s}}/2$) and the hard region
($E_{6}>\delta_{s}\sqrt{\hat{s}}/2$). The $\delta_{c}$ separates
hard region into the hard collinear ($HC$) region and hard
noncollinear ($\overline{HC}$) region. The criterion for separating
the HC region is described as follows: the region for real
gluon/light-(anti)quark emission with $\hat{s}_{16}$ (or
$\hat{s}_{15}$, $\hat{s}_{25}$, $\hat{s}_{26}$, $\hat{s}_{56}$)
$< \delta_{c}\hat{s}$ (where $\hat{s}_{ij}=(p_{i}+p_{j})^{2}$) is
called the $HC$ region. Otherwise it is called the $\overline{HC}$
region. Then the cross section for each of the real emission partonic
processes (\ref{eq-n}) can be written as
\begin{equation}
\label{siggama2} \hat{\sigma}^{R}=\hat{\sigma}^{S}+\hat{\sigma}^{H}
=\hat{\sigma}^{S}+\hat{\sigma}^{HC}+
\hat{\sigma}^{\overline{HC}}.
\end{equation}

\par
After combining the renormalized virtual corrections with the
contributions of the real gluon/light-quark emission processes and
the PDF counterterms $\delta G_{q(g)/P}$ together, the UV and IR
singularities are exactly vanished. These cancelations can be
verified numerically in our numerical calculations.

\vskip 5mm
\section{Numerical results and discussions  }
\par
In this section we present and discuss the numerical results for the
LO and QCD NLO corrected observables for the $pp \to Z^0H^0j+X$
process at the early ($\sqrt{s}=7~TeV$) and future
($\sqrt{s}=14~TeV$) LHC. We take the CTEQ6L1 PDFs with a one-loop
running $\alpha_{s}$ in the LO calculations and the CTEQ6M PDFs with
a two-loop running $\alpha_{s}$ in the NLO calculations, separately
\cite {AA3,cteq2}. The number of the active flavors is taken as
$N_{f}=5$ and the QCD parameters are set as $\Lambda^{LO}_{5}
=165~MeV$ and $\Lambda^{\overline{MS}}_{5} =226~MeV$ for the LO and
NLO calculations, respectively. For simplicity we define the
factorization scale and the renormalization scale being equal
(i.e.,$\mu=\mu_{f}=\mu_{r}$) and take
$\mu=\mu_0=\frac{1}{2}(m_{H}+m_{Z})$ by default unless otherwise
stated. The CKM matrix is set as a unit matrix and the weak mixing
angle is obtained by $c_{W}^{2}=m_{W}^{2}/m_{Z}^{2}$. We neglect
up-, down-, charm-, strange- and bottom-quark masses (i.e.,
$m_u=m_d=m_c=m_s=m_b=0~GeV$). Throughout our calculations, we take
$m_{H}=120~GeV$ by default and the other input parameters are chosen
in accordance with \cite{hepdata},
\begin{equation}
\begin{array}{lll}  \label{input1}
\alpha(m_Z^2)^{-1}=127.916~GeV,~~ m_W=80.399~GeV,~~m_Z=91.1876~GeV,~~ m_t=172.0~GeV.
\end{array}
\end{equation}

\par
As we know that the final state of the $pp \to Z^0H^0j+X$ process
contains only one jet at the LO, while it involves both one-jet
and two-jet events in final state up to the QCD NLO. The NLO
virtual correction, the real soft emission and hard collinear
emission processes include one-jet events, while the real hard
noncollinear gluon/light-(anti)quark emission processes include
two-jets events. In our calculations we apply the jet
recombination procedure of Ref.\cite{jet} in the definition of the
tagged hard jet with $R=1$. That means when two proto-jets in the
final state satisfy the constraint of $\sqrt{\Delta y^2 + \Delta
\phi^2} < R \equiv 1$, where $\Delta y$ and $\Delta \phi$ are the
differences of pseudo-rapidity and azimuthal angle between the two
proto-jets, we merge them into one new 'jet' and call it as
one-jet event, the new 'jet' four-momentum is defined as
$p_{ij,\mu}=p_{i,\mu}+p_{j,\mu}$. After applying the jet
recombination procedure to the proto-jet events of the $pp \to
Z^0H^0j+X$ up to the QCD NLO, we obtain the one-jet events and
two-jet events. As shown in Ref.\cite{Dixon}, the QCD NLO
corrections to the total cross sections at hadron colliders are
generally very large and could destroy the convergence of
perturbative description, due to the fact that the contributions
from NLO real emission subprocesses are taken into account. In
order to know how to get the modest corrections and reduce the
scale uncertainty, we adopt both the inclusive and exclusive
two-jet event selection schemes for comparison. In the following
we present detailed criterions for selecting the one-jet and
two-jet events.

\par
(1) For the one-jet events, we collect the events with the
constraint on the jet as $p_{T}^{(j)}>p_{T,j}^{cut}=50~GeV$.

\par
(2) For the two-jet events, we treat the two jets either inclusively
or exclusively. The inclusive and exclusive two-jet event selection
schemes are declared as follows:
\begin{itemize}
\item In the inclusive two-jet event selection scheme (Scheme I) both
one- and two-jet events are included, and the constraint of
$p_{T}^{(j)}> p_{T,j}^{cut}=50~GeV$ applied on the leading-jet but
not on the second-jet, where the leading-jet and the second-jet
are characterized by $p_T^{{\rm leading-jet}} > p_T^{\rm
second-jet}$.

\item In the exclusive two-jet event selection scheme (Scheme II),
the one-jet events with $p_{T}^{(j)}>p_{T,j}^{cut}=50~GeV$ are
accepted, while the two-jet events with $p_T^{\rm second-jet} >
p_{T,j}^{cut}=50~GeV$ are rejected \cite{Uwer1,Uwer2,Uwer3,44}.
\end{itemize}

\par
The verifications of the correctness of our calculations are made in
the following ways:
\par
(1) The LO calculations for the process $pp \to ug \to Z^0H^0u+X$
with $\sqrt{s}=14~TeV$ are performed by using the
FeynArts3.5/FormCalc6.0 packages and CompHEP-4.4p3 program
\cite{CompHEP}, and applying the Feynman and unitary gauges,
separately. All the results are in good agreement within the
statistic errors.

\par
(2) The virtual correction and the real
gluon/light-(anti)quark emission correction to the $pp \to Z^0H^0j+X$ process
at the LHC were evaluated twice independently based on different
codes, and they yield results in mutual agreement.

\par
(3) The UV and IR safeties are verified numerically after combining
all the NLO QCD contributions.

\par
(4) The independence of the NLO QCD correction to the \ppZHj process
on the soft cutoff $\delta_s$ and collinear cutoff $\delta_c$ is
proofed. There we apply the Scheme I for two-jet event selection,
and take $\sqrt{s}=14~TeV$, $\mu=\mu_0$, $m_H=120~GeV$ and $\delta_c
= \delta_s/50$. We find the total NLO QCD correction
$\Delta\sigma_{NLO}$ does not depend on the arbitrarily chosen value
of $\delta_s$ and $\delta_c$ within the calculation errors. In the
following numerical calculations, we fix $\delta_s=5\times10^{-4}$
and $\delta_c=\delta_s/50$.

\par
\begin{figure}[htbp]
\begin{center}
\includegraphics[scale=0.7]{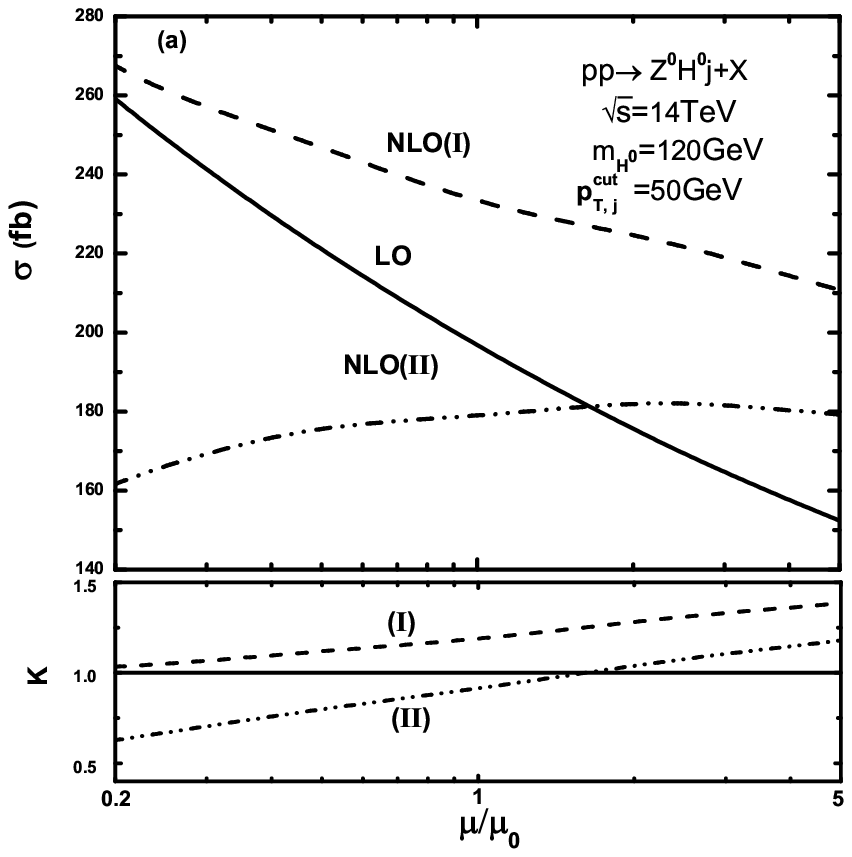}
\includegraphics[scale=0.7]{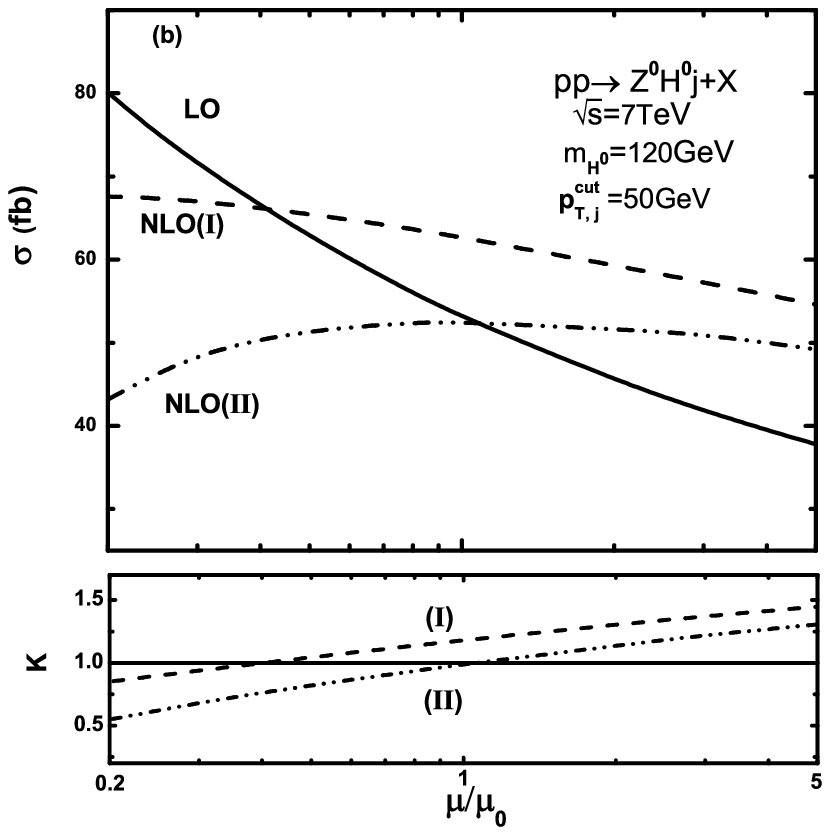}
\caption{\label{fig3} The LO, NLO QCD corrected cross sections and
the corresponding K-factor ($K(\mu)\equiv \sigma_{NLO}(\mu)/
\sigma_{LO}(\mu)$) versus the factorization/renormalization scale
$\mu$ for the \ppZHj process at the LHC by adopting the inclusive
scheme (Scheme I) and exclusive scheme (Scheme II), separately. (a)
at the $\sqrt{s}=14~TeV$ LHC. (b) at the $\sqrt{s}=7~TeV$ LHC. }
\end{center}
\end{figure}

In Figs.\ref{fig3}(a,b) we depict the dependence of the LO, NLO
QCD corrected cross sections and the corresponding K-factor for
the \ppZHj process on the renormalization/factorization scale
($\mu$) at the $\sqrt{s}=14~TeV$ and $\sqrt{s}=7~TeV$ LHC, by
adopting the inclusive and exclusive event selection schemes,
separately. Figs.\ref{fig3}(a) and (b) show that the LO curves go
down quickly with the scale running from $ 0.2~\mu_{0}$ to $5
\mu_{0}$, while the renormalization/factorization scale dependence
is obviously reduced by the NLO QCD corrections in both inclusive
and exclusive event selection schemes. The scale uncertainty is
defined as $\eta = \frac{max\{\sigma(\mu)\} -
min\{\sigma(\mu)\}}{max\{\sigma(\mu)\} + min\{\sigma(\mu)\}}$,
where $\mu \in [0.2 \mu_0,~5 \mu_0]$. We can find that at the
$\sqrt{s}=14~TeV$ LHC with the inclusive event selection scheme
(Scheme I), the scale uncertainty $\eta$ is reduced from $25.9\%$
(LO) to $11.7\%$ (NLO) as shown in Fig.\ref{fig3}(a). At the
$\sqrt{s}=7~TeV$ LHC with the inclusive event selection scheme,
the scale uncertainty is reduced from $35.8\%$ (LO) to $10.7\%$
(NLO) as shown in Fig.\ref{fig3}(b). Figs.\ref{fig3}(a,b) show
also when we apply the exclusive event selection scheme (Scheme
II), the NLO QCD corrections can reduce the scale uncertainty from
$35.8\%$ (LO) to $5.9\%$ and $9.7\%$ (NLO) with $\sqrt{s} =
14~TeV$ and $7~TeV$, respectively. We can also see that the
K-factor varies between $0.996~(0.69)$ and $1.46~(1.55)$ at
$\sqrt{s}=14~TeV~(7~TeV)$ by adopting the inclusive event
selection scheme, and between $0.50~(0.35)$ and $1.27~(1.42)$ at
$\sqrt{s}=14~TeV~(7~TeV)$ by adopting the exclusive event
selection scheme in the plotted range of $\mu$. Comparing the
curves in Figs.\ref{fig3}(a,b), we can see the different line
shapes for the two curves labeled with 'NLO(I)', which are for the
cross sections obtained by adopting the inclusive scheme (Scheme
I) at the $\sqrt{s}=14~TeV$ and $\sqrt{s}=7~TeV$ LHC,
respectively. That difference comes from that the NLO QCD
correction component ($\sigma^{\overline{HC}}$) of the
non-collinear real hard emission correction at the
$\sqrt{s}=14~TeV$ LHC is proportional to $\alpha_s^2(\mu)$, and
its contribution to integrated cross section is much larger than
that at the $\sqrt{s}=7~TeV$ LHC. Therefore, the line shape for
$\sqrt{s}=14~TeV$ in Fig.\ref{fig3}(a) shows the cross section
adopting the inclusive scheme is going down with the increment of
the scale.

\par
In Table \ref{tab1}, we list the numerical results for the LO, NLO
QCD corrected cross sections and the corresponding K-factor
($K\equiv\frac{\sigma_{NLO}}{\sigma_{LO}}$) for the process \ppZHj
at the LHC by applying the Scheme I and Scheme II with
$\mu=\mu_{0}$, $\mu=\mu_1\equiv \frac{1}{2}(E_T^Z+
E_T^H+\sum_{j}E_T^j)$ and $\sqrt{s}=14~TeV,~7~TeV$, separately.
From this table, we can see that the LO ($\sigma_{LO}$) and NLO
QCD corrected cross sections by adopting the inclusive and
exclusive event selection schemes ($\sigma_{NLO}^{(I)}$,
$\sigma_{NLO}^{(II)}$) are all sensitive to the scale choices, and
the difference between the NLO QCD corrected cross sections
obtained by taking $\mu=\mu_0$ and $\mu=\mu_1$ respectively, is
about $3\%$, which is much smaller than the discrepancy of LO
cross sections with these two scale choices (about $10\%$).
\begin{table}
\begin{center}
\begin{tabular}{|c|c|c|c|c|c|c|}
\hline $pp \to Z^0H^0j+X$ &$\mu(GeV)$ & $\sigma_{LO}(fb)$ &
$\sigma_{NLO}^{(I)}(fb)$ & $\sigma_{NLO}^{(II)}(fb)$ & $K^{(I)}$ & $K^{(II)}$ \\
\hline        $\sqrt{s}=14~TeV$   & $\mu_{0}$ &196.25(1)  & 232.2(2)  & 178.7(2) & 1.1832 & 0.9106 \\
\cline{2-7}                       & $\mu_{1}$ & 175.03(1) & 226.4(6)  & 184.9(6)  & 1.2933 & 1.0564 \\
\hline        $\sqrt{s}=7~TeV$    &$\mu_{0}$  & 53.116(3) & 62.67(7)  & 52.50(7) & 1.1799 & 0.9884  \\
\cline{2-7}                       &$\mu_{1}$  & 46.022(3) & 60.86(5)  & 53.71(5) & 1.3225& 1.1670\\
\hline
\end{tabular}
\end{center}
\begin{center}
\begin{minipage}{14cm}
\caption{\label{tab1} The numerical results for the LO, NLO QCD
corrected cross sections and the corresponding K-factor
($K\equiv\frac{\sigma_{NLO}}{\sigma_{LO}}$) with
$p_{T,j}^{cut}=50~GeV$, $m_{H}=120~GeV$ for the $pp \to Z^0H^0j+X$
process at the LHC by taking
$\mu=\mu_0=\frac{1}{2}(m_{H}+m_{Z})=105.594~GeV$ and
$\mu=\mu_1=\frac{1}{2}(E_{T}^{Z}+E_{T}^{H}+\sum
_{jet}E_{T}^{jet})$ and applying the Scheme I and Scheme II event
selection schemes, respectively. }
\end{minipage}
\end{center}
\end{table}

\par
\begin{figure}
\centering
\includegraphics[scale=0.8]{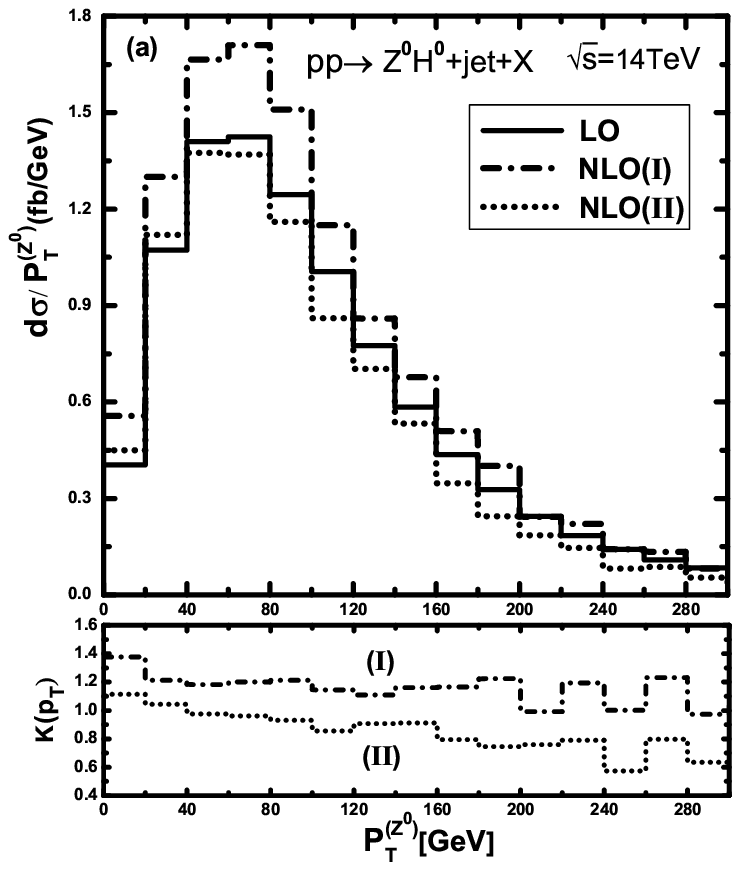}
\includegraphics[scale=0.8]{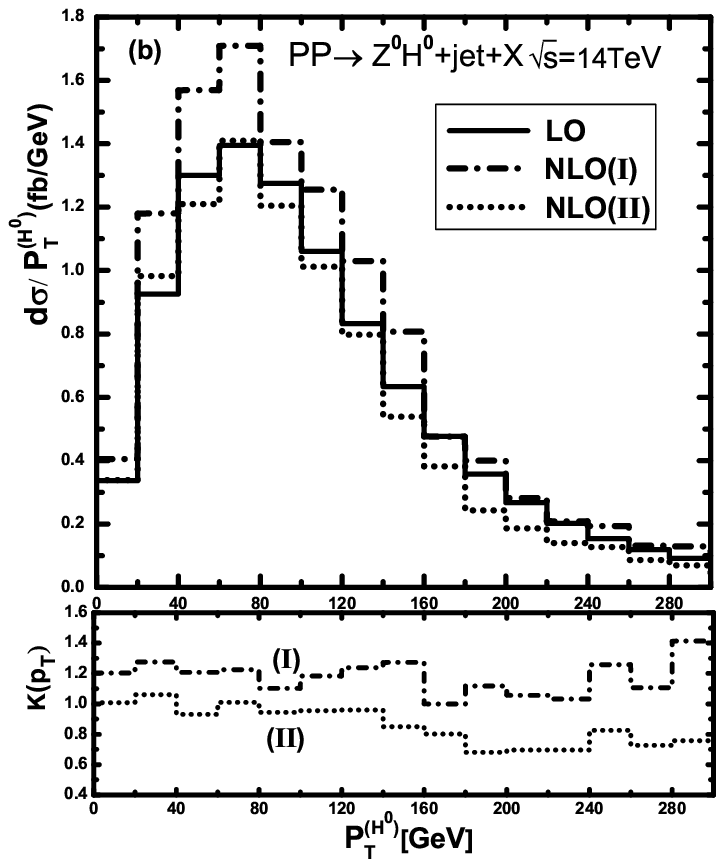}
\includegraphics[scale=0.8]{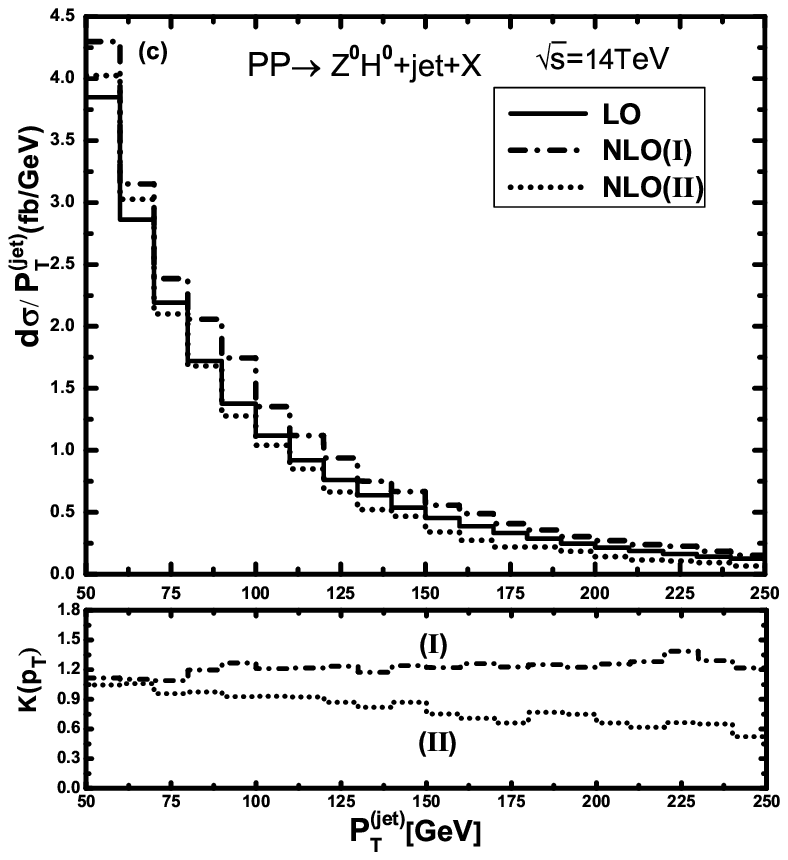}
\caption{\label{fig4} The LO, NLO QCD corrected distributions of
the transverse momenta of the final particles and the
corresponding K-factors ($K(p_T)\equiv \frac{d
\sigma_{NLO}}{dp_T}/ \frac{d \sigma_{LO}}{dp_T}$) for the $pp \to
Z^0H^0j+X$ process at the LHC with $\mu=\mu_0$ and
$\sqrt{s}=14~TeV$. (a) $Z^0$-boson, (b) Higgs-boson, (c) final
leading-jet. }
\end{figure}
\begin{figure}
\centering
\includegraphics[scale=0.75]{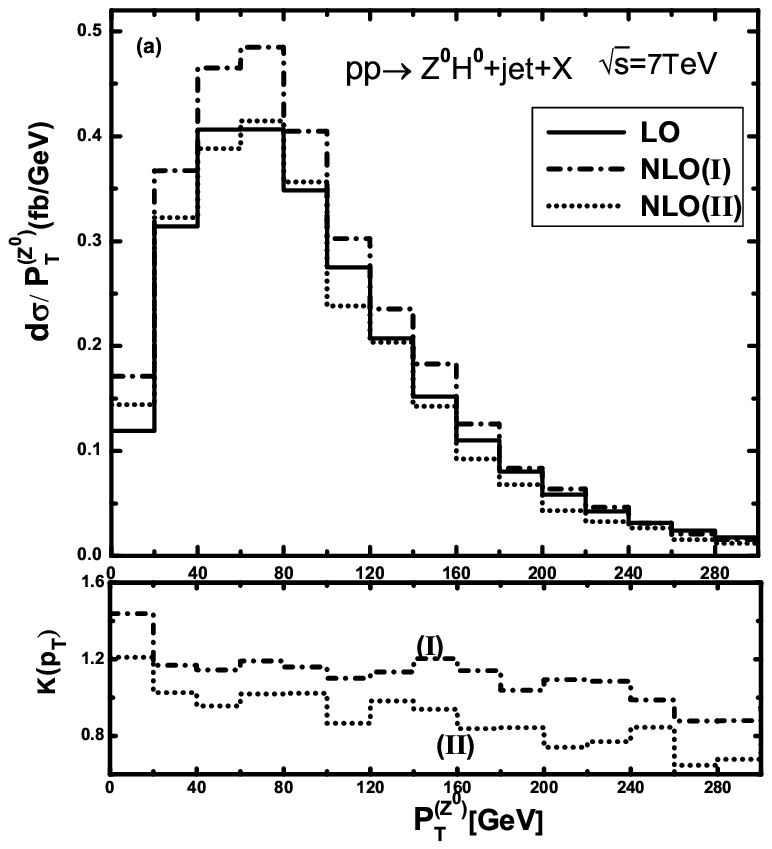}
\includegraphics[scale=0.75]{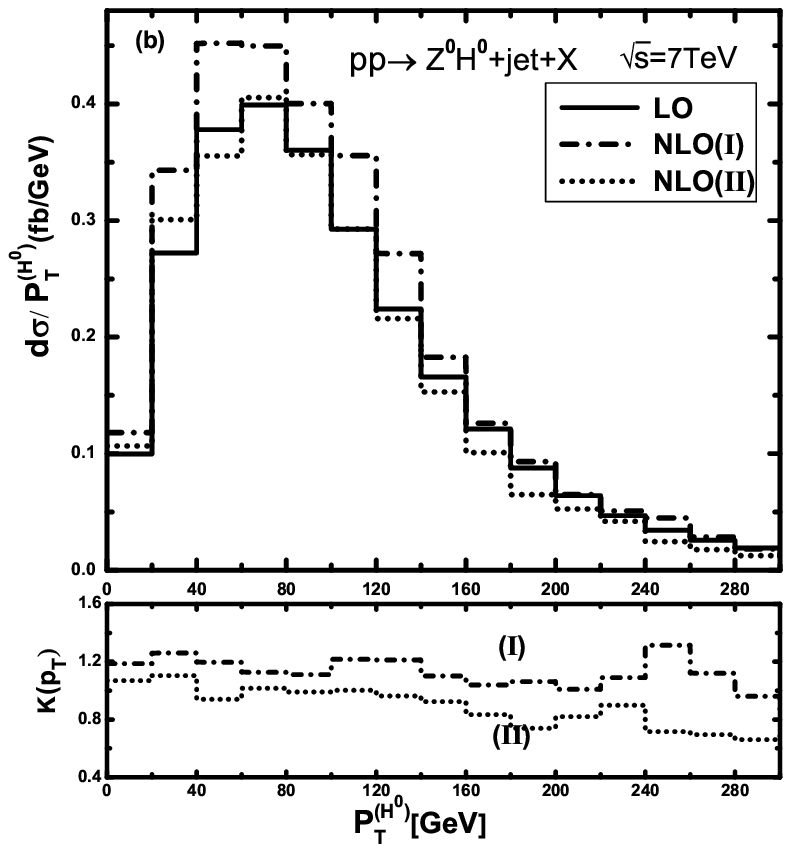}
\includegraphics[scale=0.75]{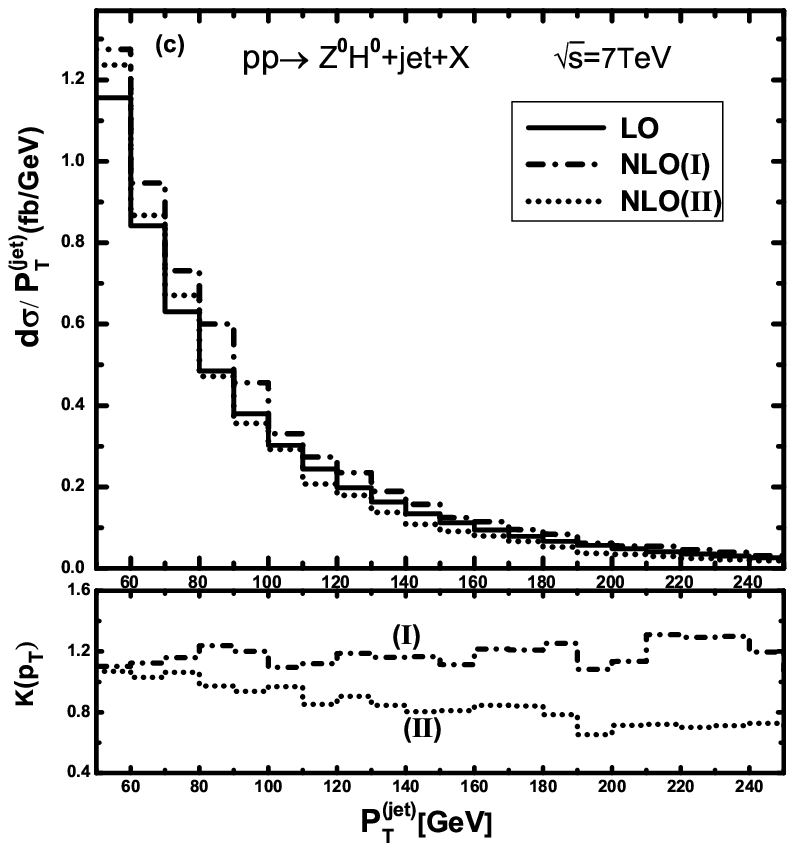}
\caption{\label{fig5} The LO, NLO QCD corrected distributions of
the transverse momenta of the final particles and the
corresponding K-factors ($K(p_T)\equiv \frac{d
\sigma_{NLO}}{dp_T}/ \frac{d \sigma_{LO}}{dp_T}$) for the $pp \to
Z^0H^0j+X$ process at the LHC with $\mu=\mu_0$ and
$\sqrt{s}=7~TeV$. (a) $Z^0$-boson, (b) Higgs-boson, (c) final
leading-jet. }
\end{figure}

In Figs.\ref{fig4}(a,b,c), we present the LO and NLO QCD corrected
distributions of the transverse momenta of the final $Z^0$-boson,
$H^0$-boson and leading-jet ($p_T^{(Z^0)}, p_T^{(H^0)},
p_T^{(j)}$) for the process \ppZHj with $\mu=\mu_0$ at the
$\sqrt{s}=14~TeV$ LHC, by adopting the inclusive and exclusive
event selection schemes, respectively. The corresponding K-factors
($K(p_T)\equiv \frac{d\sigma_{NLO}}{dp_T}/
\frac{d\sigma_{LO}}{dp_T}$) are also shown in these figures. The
analogous plots for the \ppZHj process with $\mu=\mu_0$ at the
$\sqrt{s}=7~TeV$ LHC are presented in Figs.\ref{fig5}(a,b,c). In
these 6 figures, the curves by applying the inclusive event
selection scheme demonstrate that the NLO QCD corrections
generally significantly enhance the LO differential cross sections
for the \ppZHj process, especially in the range of $p_T^{(Z^0)},
p_T^{(H^0)}, p_T^{(j)} < 150~GeV$, while the curves by adopting
the exclusive event selection scheme show that the NLO QCD
corrections increase the LO differential cross sections only in
the low transverse momentum ranges of
$p_{T}^{(Z^0)},p_{T}^{(H^0)}<40~GeV$ and $p_{T}^{(j)}< 70~GeV$. We
also see from Figs.\ref{fig4}(a,b,c) and Figs.\ref{fig5}(a,b,c)
that the NLO QCD correction does not obviously change the shape of
the LO distribution of the final particles by using both the
inclusive and exclusive event selection schemes.

\par
\begin{figure}
\centering
\includegraphics[scale=0.8]{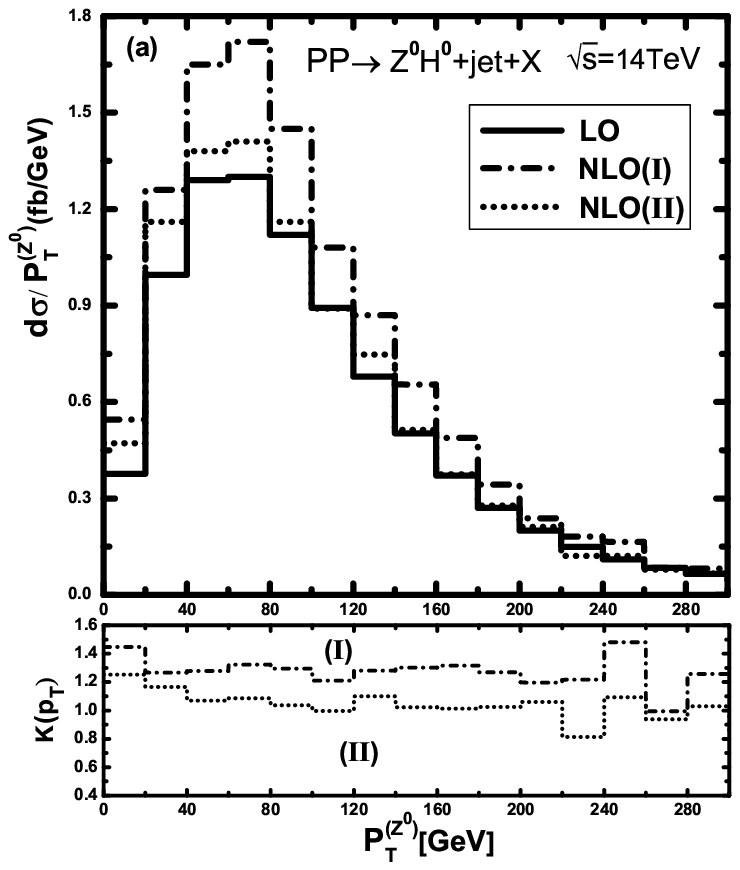}
\includegraphics[scale=0.8]{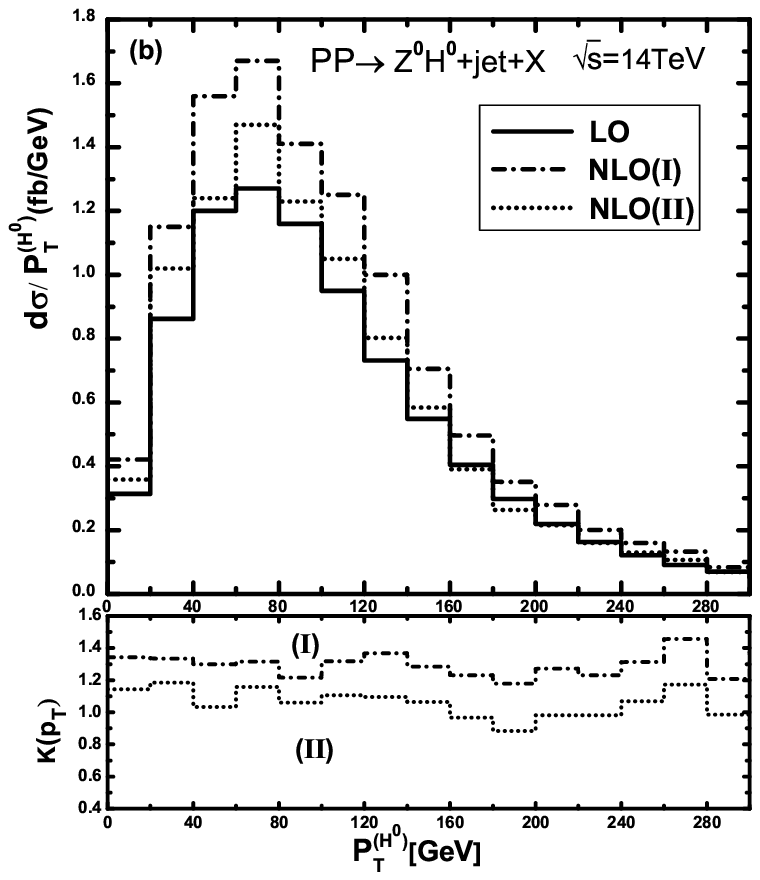}
\includegraphics[scale=0.8]{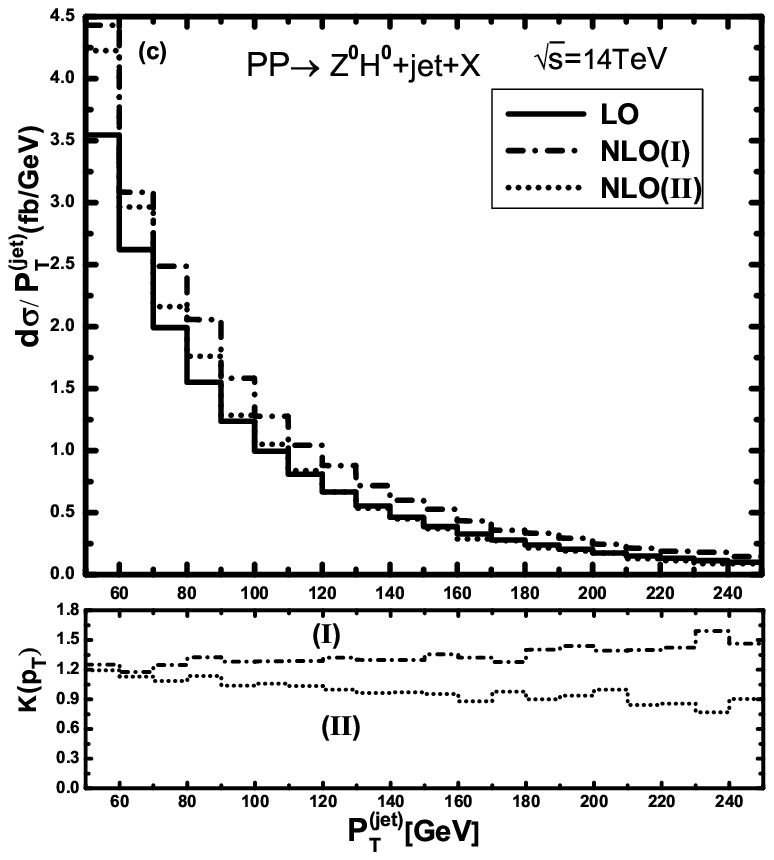}
\caption{\label{fig6} The LO, NLO QCD corrected distributions of
the transverse momenta of the final particles and the
corresponding K-factors ($K(p_T)\equiv \frac{d
\sigma_{NLO}}{dp_T}/ \frac{d \sigma_{LO}}{dp_T}$) for the $pp \to
Z^0H^0j+X$ process at the LHC with
$\mu=\mu_1=\frac{1}{2}(E_{T}^{Z}+E_{T}^{H}+ \sum
_{jet}E_{T}^{jet})$ and $\sqrt{s}=14~TeV$. (a) $Z^0$-boson, (b)
Higgs-boson, (c) final leading-jet. }
\end{figure}
\begin{figure}
\centering
\includegraphics[scale=0.75]{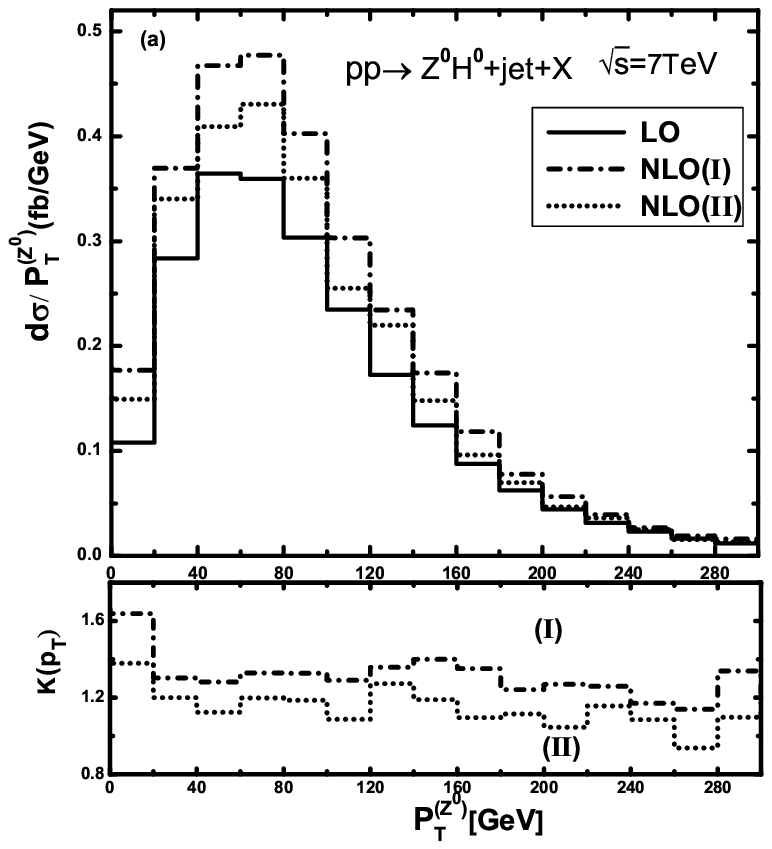}
\includegraphics[scale=0.75]{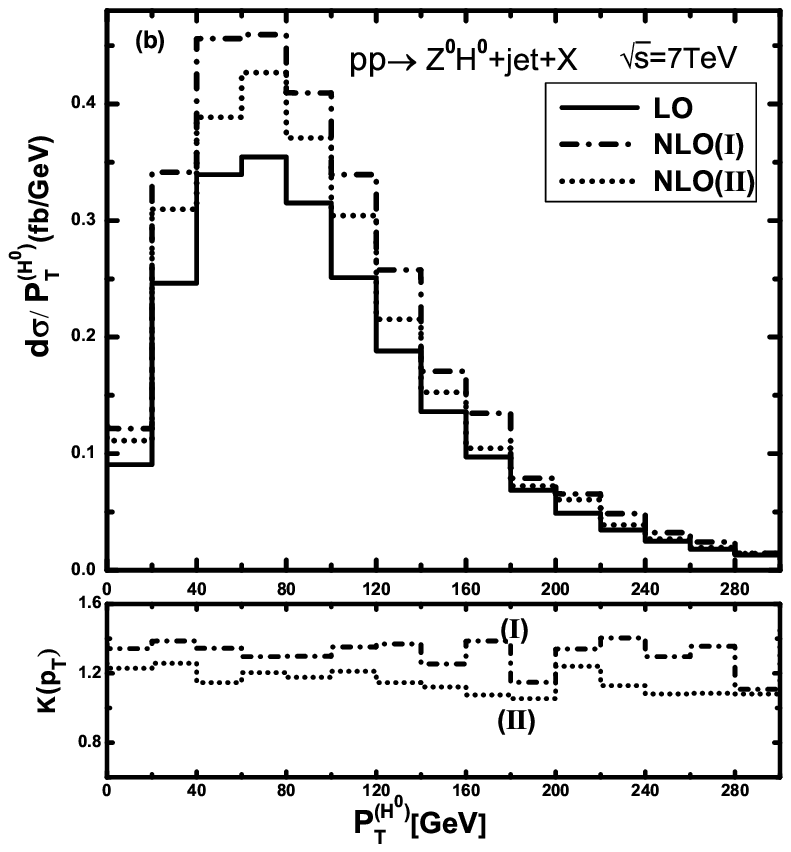}
\includegraphics[scale=0.75]{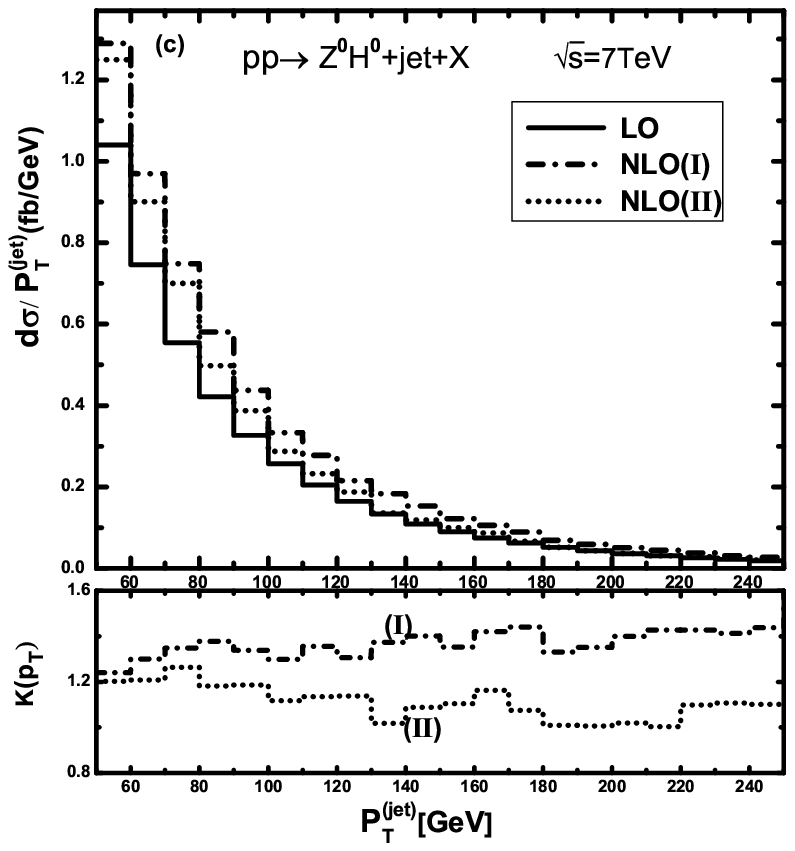}
\caption{\label{fig7} The LO, NLO QCD corrected distributions of
the transverse momenta of the final particles and the
corresponding K-factors ($K(p_T)\equiv \frac{d
\sigma_{NLO}}{dp_T}/ \frac{d \sigma_{LO}}{dp_T}$) for the $pp \to
Z^0H^0j+X$ process at the LHC with
$\mu=\mu_1=\frac{1}{2}(E_{T}^{Z}+E_{T}^{H}+ \sum
_{jet}E_{T}^{jet})$ and $ \sqrt{s}=7~TeV$. (a) $Z^0$-boson, (b)
Higgs-boson, (c) final leading-jet. }
\end{figure}

For the further analysis of the uncertainty due to the scale
variation, we take another scale choice, i.e., by implying the
scale $\mu_1 =\frac{1}{2}(E_{T}^{Z}+E_{T}^{H}+ \sum
_{jet}E_{T}^{jet})$ which is relevant to the transverse energies
of final particles as the factorization/renormalization scale
$\mu=\mu_f=\mu_r$ \cite{Berger}, and compare with the case of
$\mu=\mu_0$. The LO and NLO QCD corrected transverse momentum
distributions of final $Z^0$-boson, $H^0$-boson and leading-jet,
and their corresponding K-factors with $\mu=\mu_1$ at the
$\sqrt{s}=14~TeV$ and $\sqrt{s}=7~TeV$ LHC, are plotted in
Figs.\ref{fig6}(a,b,c) and Figs.\ref{fig7}(a,b,c), separately. In
comparison with the corresponding curves in Figs.\ref{fig4}(a,b,c)
and Figs.\ref{fig5}(a,b,c), we can see the transverse momentum
distribution curves for two scale choice are quite similar, but
the curves of $K(p_T)$-factor for $\mu=\mu_1$ are somewhat more
stable than the corresponding ones for $\mu=\mu_0$. It seems that
the scale choice of the phase space dependent scale $\mu=\mu_1$ is
more appropriate than the fixed one $\mu=\mu_0$ for the
$Z^0H^0+jet$ production process at the LHC.

\vskip 5mm
\section{ Summary }
\par
In this work, we present the a full treatment of the
next-to-leading order QCD corrections to the observables of the
$pp \to Z^0H^0j+X$ at the early ($\sqrt{s}=7~TeV$) and future
($\sqrt{s}=14~TeV$) LHC. We investigate the dependence of the LO
and the NLO QCD corrected cross sections on the
renormalization/factorization scale, and study the LO and NLO QCD
corrected distributions of the transverse momenta for the final
particles ($Z^0,~H^0$, leading-jet) by adopting the inclusive and
exclusive event selection schemes. By taking
$\mu=\mu_{0}=(m_Z+m_H)/2$, $m_{H}=120~GeV $ and the jet constraint
of $p_{T}^{(j)} > p_{T,j}^{cut}=50~GeV$, we find the K-factors for
the total cross section have the values of $1.183~(0.988)$ at the
$\sqrt{s}=14~TeV$ ($\sqrt{s}=7~TeV$) LHC by adopting the inclusive
and exclusive schemes, respectively. Our numerical results also
show that the NLO QCD corrections obviously modify the LO
integrated and differential cross sections, and significantly
reduce the scale uncertainty of the LO cross section by adopting
either the inclusive or the exclusive schemes. We find also that a
scale choice like $\mu=\mu_1=\frac{1}{2}(E_{T}^{Z}+E_{T}^{H}+ \sum
_{j}E_{T}^{jet})$ seems to be more appropriate than the fixed
scale $\mu=\mu_0$.

\vskip 5mm
\par
\acknowledgments{This work was supported in
part by the National Natural Science Foundation of China (Contract
No.10875112, No.11075150, No.11005101), and the Specialized Research
Fund for the Doctoral Program of Higher Education (Contract
No.20093402110030).}

\appendix
\renewcommand{\theequation}{\thesection.\arabic{equation}}
\section{Some expressions of the amplitudes at LO and NLO }

The partonic processes of the $Z^{0}H^0+jet$ production at the LHC
are related to the amplitudes of partonic processes $q(p_1)\bar q(p_2)
\to Z^0(p_3)H^0(p_4)g(p_5)$ and $q(\bar q)(p_1)g(p_2) \to Z^0(p_3)H^0(p_4)q(\bar q)(p_5)$.
The summations of the squared amplitude over all spins and colors for the
$q(p_1)\bar q(p_2) \to Z^0(p_3)H^0(p_4)g(p_5)$,
$q(p_1)g(p_2) \to Z^0(p_3)H^0(p_4)q(p_5)$ and $\bar q(p_1)g(p_2) \to
Z^0(p_3)H^0(p_4)\bar q(p_5)$ partonic processes at the LO are expressed separately as
\begin{eqnarray}
\sum^{spin}_{color}|{\cal M}_{LO}^{q\bar{q}}|^2&=&\frac{4g_s^2e^4 m_W^2(32 s_W^2-24 s_W^2+9)}{9c_W^6s_W^4m_Z^2
[(p_3+p_4)^2-m_Z^2]^2(p_2-p_5)^4}[m_Z^2(2p_1\cdot p_5\ p_2\cdot p_5
-m_z^2p_1\cdot p_2) \nb\\
&&+2p_1\cdot p_3(2p_3\cdot p_5\ p_2\cdot p_5)]-\frac{4g_s^2e^4 m_W^2(32 s_W^2-24 s_W^2+9)}{9c_W^6s_W^4m_Z^2
[(p_3+p_4)^2-m_Z^2]^2(p_2-p_3-p_4)^4} \nb\\
&&\{2p_1\cdot p_4[(2p_2\cdot p_3)^2-2p_3\cdot p_4\ p_2\cdot p_3-m_Z^2p_2\cdot p_4]
+2p_1\cdot p_4[-m_Z^2p_2\cdot p_4 \nb\\
&&+p_2\cdot p_4(m_H^2-2m_Z^2-2p_2\cdot p_4)]+p_1\cdot p_2[-4(p_2\cdot p_3)^2+
4p_3\cdot p_4\ p_2\cdot p_3+m_Z^2(m_Z^2 \nb\\
&&+m_H^2+2p_3\cdot p_4)]\}+\frac{4g_s^2e^4 m_W^2(32 s_W^2-24 s_W^2+
9)}{9c_W^6s_W^4m_Z^2[(p_3+p_4)^2-m_Z^2]^2(p_2-p_5)^2(p_2-p_3-p_4)^2} \nb\\
&&\{m_Z^2(p_1\cdot p_3\ p_2\cdot p_5-p_1\cdot p_4 p_2\cdot p_4)+p_1\cdot p_5[
m_Z^2p_2\cdot p_4-(2p_3\cdot p_4+m_Z^2)p_2\cdot p_3 \nb\\
&&+2(p_2\cdot p_3)^2]+p_2\cdot p_2[m_Z^2(2p_2\cdot p_4+4p_2\cdot p_5-3p_3\cdot
p_5-3p_4\cdot p_5)+2(p_3\cdot p_4 \nb\\
&&-p_3\cdot p_5+2m_Z^2)p_2\cdot p_3]-2p_1\cdot p_4(p_2\cdot p_3)^2+2p_1\cdot
p_3\ p_2\cdot p_3\ p_2\cdot p_4 \nb\\
&&-2p_1\cdot p_3\ p_2\cdot p_3\ p_2\cdot p_5+2p_3\cdot p_4\ p_1\cdot p_3\ p_2\cdot p_5
+2p_3\cdot p_5\ p_1\cdot p_4\ p_2\cdot p_3 \nb\\
&&-2p_3\cdot p_5\ p_1\cdot p_3\ p_2\cdot p_4 \},\\
\sum^{spin}_{color}|{\cal M}_{LO}^{qg}|^2&=&|M_{LO}^{q\bar q}|^2\{p_2\leftrightarrow -p_5\},  \\
\sum^{spin}_{color}|{\cal M}_{LO}^{\bar qg}|^2&=&|M_{LO}^{q\bar
q}|^2\{p_1\to -p_5,p_5\to -p_2, p_2\to p_1\},
\end{eqnarray}
where the summations are taken over spins and colors of all
particles involved.

\par
The real emission partonic processes for the $Z^{0}H^0+jet$
production at the LHC are related to the amplitudes of $0 \to
q\bar qggZ^0H^0$ and $0\to q\bar q q'\bar q'Z^0H^0$ by crossing
symmetry. The amplitude of the $0\to q(p_1,c_1)\bar
q(p_2,c_2)g(p_3,c_3) g(p_4,c_4)Z^0(p_5)H^0(p_6)$ partonic process
at the tree-level is expressed as below.
\begin{eqnarray}
&&{\cal M}(0\to q\bar q gg Z^0H^0)= \nb \\
&&\frac{T_{c_1,x}^{c_3}T_{x,c_2}^{c_4}\varepsilon^{*}_{\mu_1}(p_5)
\varepsilon^{*}_{\mu_2}(p_3)\varepsilon^{*}_{\mu_3}(p_4)C_{ZZH}}
{(p_1+p_3)^2[(p_5+p_6)^2-m_Z^2](p_2+p_4)^2}
\bar u(p_1)C_{q\bar qg}^{\mu_2}(\slashed{p_1}+\slashed{p_3})
C_{q\bar qZ}^{\mu_1}(-\slashed{p_2}-\slashed{p_4})C_{q\bar qg}^{\mu_3}v(-p_2)\nb\\
&+&\frac{T_{c_1,x}^{c_4}T_{x,c_2}^{c_3}\varepsilon^{*}_{\mu_1}
(p_5)\varepsilon^{*}_{\mu_2}(p_3)\varepsilon^{*}_{\mu_3}(p_4)C_{ZZH}}
{(p_1+p_4)^2[(p_5+p_6)^2-m_Z^2](p_2+p_3)^2}
\bar u(p_1)C_{q\bar qg}^{\mu_2}(\slashed{p_1}+\slashed{p_4})
C_{q\bar qZ}^{\mu_1}(-\slashed{p_2}-\slashed{p_3})C_{q\bar qg}^{\mu_3}v(-p_2)\nb\\
&+&\frac{T_{c_1,x}^{c_3}T_{x,c_2}^{c_4}\varepsilon^{*}_{\mu_1}(p_5)
\varepsilon^{*}_{\mu_2}(p_3)\varepsilon^{*}_{\mu_3}(p_4)C_{ZZH}}
{(p_1+p_3)^2[(p_5+p_6)^2-m_Z^2](p_1+p_3+p_4)^2}
\bar u(p_1)C_{q\bar qg}^{\mu_2}(\slashed{p_1}+\slashed{p_3})
C_{q\bar qg}^{\mu_3}(\slashed{p_1}+\slashed{p_3}+\slashed{p_4})
C_{q\bar qZ}^{\mu_1}v(-p_2)\nb\\
&+&\frac{T_{c_1,x}^{c_4}T_{x,c_2}^{c_3}\varepsilon^{*}_{\mu_1}
(p_5)\varepsilon^{*}_{\mu_2}(p_3)\varepsilon^{*}_{\mu_3}(p_4)C_{ZZH}}
{(p_1+p_4)^2[(p_5+p_6)^2-m_Z^2](p_1+p_3+p_4)^2}
\bar u(p_1)C_{q\bar qg}^{\mu_2}(\slashed{p_1}+\slashed{p_4})
C_{q\bar qg}^{\mu_3}(\slashed{p_1}+\slashed{p_3}+\slashed{p_4})
C_{q\bar qZ}^{\mu_1}v(-p_2)\nb\\
&+&\frac{T_{c_1,x}^{c_3}T_{x,c_2}^{c_4}\varepsilon^{*}_{\mu_1}(p_5)
\varepsilon^{*}_{\mu_2}(p_3)\varepsilon^{*}_{\mu_3}(p_4)C_{ZZH}}
{(p_1+p_5+p_6)^2[(p_5+p_6)^2-m_Z^2](p_2+p_4)^2}
\bar u(p_1)C_{q\bar qZ}^{\mu_1}(\slashed{p_1}+\slashed{p_5}+\slashed{p_6})
C_{q\bar qg}^{\mu_2}(-\slashed{p_2}-\slashed{p_4})C_{q\bar qg}^{\mu_3}v(-p_2)\nb\\
&+&\frac{T_{c_1,x}^{c_4}T_{x,c_2}^{c_3}\varepsilon^{*}_{\mu_1}
(p_5)\varepsilon^{*}_{\mu_2}(p_3)\varepsilon^{*}_{\mu_3}(p_4)C_{ZZH}}
{(p_1+p_5+p_6)^2[(p_5+p_6)^2-m_Z^2](p_2+p_3)^2}
\bar u(p_1)C_{q\bar qZ}^{\mu_1}(\slashed{p_1}+\slashed{p_5}+\slashed{p_6})
C_{q\bar qg}^{\mu_2}(-\slashed{p_2}-\slashed{p_3})C_{q\bar qg}^{\mu_3}v(-p_2)\nb\\
&+&\frac{T_{c_1c_2}^{x}f^{xc_3c_4}\varepsilon^{*}_{\mu_1}
(p_5)\varepsilon^{*}_{\mu_2}(p_3)\varepsilon^{*}_{\mu_3}(p_4)C_{ZZH}}
{[(p_5+p_6)^2-m_Z^2](p_3+p_4)^2(p_1+p_3+p_4)^2}\bar u(p_1)C_{q\bar qg}^{\nu_1}
(\slashed{p_1}+\slashed{p_3}+\slashed{p_4})C_{q\bar qZ}^{\mu_1}v(-p_2)
C_{ggg}^{\mu_2\mu_3\nu_2}g_{\nu_1\nu_2}\nb\\
&+&\frac{T_{c_1c_2}^{x}f^{xc_3c_4}\varepsilon^{*}_{\mu_1}
(p_5)\varepsilon^{*}_{\mu_2}(p_3)\varepsilon^{*}_{\mu_3}(p_4)C_{ZZH}}
{[(p_5+p_6)^2-m_Z^2](p_3+p_4)^2(p_1+p_5+p_6)^2}\bar u(p_1)C_{q\bar qZ}^{\mu_1}
(\slashed{p_1}+\slashed{p_5}+\slashed{p_6})C_{q\bar qg}^{\nu_1}v(-p_2)
C_{ggg}^{\mu_2\mu_3\nu_2}g_{\nu_1\nu_2}. \nb \\
\end{eqnarray}

\par
The amplitude of the $0\to q(p_1,c_1)\bar q(p_2,c_2) q'(p_3,c_3)
\bar q'(p_4,c_4)Z^0(p_5)H^0(p_6)$ partonic process at the
tree-level can be expressed as
\begin{eqnarray}
&&{\cal M}(0\to q\bar q q'\bar{q}'Z^0H^0)= \nb \\
&&\frac{T_{c_1c_2}^xT_{c_3c_4}^x\varepsilon^{*}_{\mu_1}
(p_5)C_{ZZH}g_{\nu_1\nu_2}}{(p_1+p_2)^2[(p_5+p_6)^2-m_Z^2](p_4+p_5+p_6)^2}
\bar u(p_1)C_{q\bar q g}^{\nu_1}v(-p_2)\bar u(p_3)C_{q'\bar q'
g}^{\nu_2}
(\slashed{p_1}+\slashed{p_2}+\slashed{p_3})\nb\\
&&C_{q'\bar q' Z}^{\mu_1}v(-p_4)
+\frac{T_{c_1c_2}^xT_{c_3c_4}^x\varepsilon^{*}_{\mu_1}
(p_5)C_{ZZH}g_{\nu_1\nu_2}}{(p_1+p_2)^2[(p_5+p_6)^2-m_Z^2](p_3+p_5+p_6)^2}
\bar u(p_1)C_{q\bar q g}^{\nu_1}v(-p_2)\bar u(p_3)\nb\\
&&C_{q'\bar q' Z}^{\mu_1}
(\slashed{p_3}+\slashed{p_5}+\slashed{p_6})C_{q'\bar q'
g}^{\nu_2}v(-p_4)
+\frac{T_{c_1c_2}^xT_{c_3c_4}^x\varepsilon^{*}_{\mu_1}
(p_5)C_{ZZH}g_{\nu_1\nu_2}}{(p_3+p_4)^2[(p_5+p_6)^2-m_Z^2](p_2+p_5+p_6)^2}
\bar u(p_3)C_{q'\bar q' g}^{\nu_1}\nb\\
&&v(-p_4)\bar u(p_1)C_{q\bar q g}^{\nu_2}
(\slashed{p_1}+\slashed{p_3}+\slashed{p_4})C_{q\bar q
Z}^{\mu_1}v(-p_2)
+\frac{T_{c_1c_2}^xT_{c_3c_4}^x\varepsilon^{*}_{\mu_1}
(p_5)C_{ZZH}g_{\nu_1\nu_2}}{(p_3+p_4)^2[(p_5+p_6)^2-m_Z^2](p_1+p_5+p_6)^2}\nb\\
&&\bar u(p_3)C_{q'\bar q' g}^{\nu_1}v(-p_4)\bar u(p_1)C_{q\bar q
Z}^{\mu_1} (\slashed{p_1}+\slashed{p_5}+\slashed{p_6})C_{q\bar q
g}^{\nu_2}v(-p_2),
\end{eqnarray}
where $p_i$ and $c_i$ are the momentum and color of the $i$-th
particle, $x$ is the color index of the gluon/quark propagator and
\begin{eqnarray}
C_{ZZH}&=&\frac{i\ e\ m_W}{c_W^2s_W},~~~~~
C_{q\bar q g}^{\mu}= -i\ g_s\gamma^{\mu}, \\
C_{q\bar q Z}^{\mu}&=&\frac{i\ e}{c_Ws_W}\left[\left(\frac{I_3}{2}-Q_qs_W^2\right)
\gamma^{\mu}-\frac{I_3}{2}\gamma^{\mu}\gamma^{5} \right], \\
C_{ggg}^{\mu_1\mu_2\nu_2}&=&g_s\left[ g^{\mu_1\mu_2}(p_3-p_4)^{\nu_1}+
g^{\mu_2\nu_1}(p_3+2p_4)^{\mu_1}+g^{\nu_1\mu_1}(-2p_3-p_4)^{\mu_2}\right].
\end{eqnarray}
In above equations $I_3^q$ and $Q_q$ are the weak isospin and the charge of quark $q$,
respectively.


\vskip 10mm
\begin{thebibliography}{99}
\bibitem{1}
    P. W. Higgs, {\it Broken Symmetries and the Masses of Gauge Bosons,
    Phys. Rev. Lett.} {\bf 13} (1964) 508.

\bibitem{AA1}
    A. Djouadi, {\it The anatomy of electroweak symmetry breaking:
    Tome I: The Higgs boson in the Standard Model
    Phys. Rept.} {\bf 457} (2008)1.

\bibitem{AA2}
    A. Djouadi, {\it The anatomy of electroweak symmetry breaking:
    Tome II: The Higgs boson in the Standard Model
    Phys. Rept.} {\bf 459} (2008)1.

\bibitem{AA3}
     J. Pumplin, D.R. Stump, J. Huston, H.L. Lai, P. Nadolsky, and W.K. Tung,
     {\it New Generation of Parton Distributions with Uncertainties from
     Global QCD Analysis, JHEP} {\bf 0207} (2002) 012, [arXiv:hep-ph/0201195].

\bibitem{2}
    G. Abbiendi.et al(the ALEPH, the DELPHI, the L3 and the OPAL, The
    LEP Working Group fot Higgs Boson Searches),{\it Search for the
    Standard Model Higgs boson at LEP, Phys. Lett. B} {\bf 565} (2003) 61.

\bibitem{mh-Fermilab}
    The CDF, D0 Collaborations, the Tevatron New Phenomena, Higgs Working Group,
    {\it Combined CDF and D0 Upper Limits on Standard Model Higgs Boson Production
    with up to $8.6~fb^{-1}$ of Data}, FERMILAB-CONF-11-354-E, [arXiv:1107.5518].

\bibitem{mh-Atlas}
    ATLAS Collaboration, [https://twiki.cern.ch/twiki/bin/view/AtlasPublic/AtlasResultsEPS2011].

\bibitem{mh-CMS}
    CMS Collaboration,{\it Combination of Higgs Searches}, CMS PAS HIG-11-022,
    [http://cms.web.cern.ch/cms/News/2011/LP11].

\bibitem{mh-LHC}
    M. Baak, M. Goebel, J. Haller, A. Hoecker, D. Ludwig, K. Moenig, M. Schott,
    J. Stelzer, {\it Updated Status of the Global Electroweak Fit and Constraints
    on New Physics}, [arXiv:1107.0975].

\bibitem{mh-Atlas-1}
    ATLAS Collaboration, ``ATLAS experiment presents latest Higgs search status'',
    http://www.atlas.ch/news/2011/status-report-dec-2011.html

\bibitem{mh-CMS-1}
    CMS Collaboration,``CMS search for the Standard Model Higgs Boson in LHC
    data from 2010 and 2011'',
    http://cms.web.cern.ch/news/cms-search-standard-model-higgs-boson-lhc-data-2010-and-2011.

\bibitem{9}
    M. Spira, A. Djouadi, D. Graudenz and P. M. Zerwas, {\it HIGGS BOSON PRODUCTION
    AT THE LHC, Nucl. Phys. B} {\bf 453}, (1995) 17, [arXiv:hep-ph/9504378].

\bibitem{10}
    C. Anastasiou, R. Boughezal and F. Petriello, {\it Mixed QCD-electroweak corrections
    to Higgs boson production in gluon fusion, JHEP} {\bf 0904} (2009) 003, [arXiv:0811.3458].

\bibitem{11}
    D. L. Rainwater, D. Zeppenfeld and K. Hagiwara, {\it Searching for $H \to \tau^+ \tau^-$
    in weak boson fusion at the LHC, Phys. Rev. D} {\bf 59} (1999) 014037, [arXiv:hep-ph/9808468].

\bibitem{12}
    W. Beenakker, S. Dittmaier, M. Kramer, B. Plumper, M. Spira and P. M. Zerwas,
    {\it NLO QCD corrections to t anti-t H production in hadron collisions,
    Nucl. Phys. B} {\bf 653} (2003) 151, [arXiv:hep-ph/0211352].

\bibitem{13}
    O. Brein, M. Ciccolini, S. Dittmaier, A. Djouadi, R. Harlander and M. Kramer,
    {\it Precision Calculations for Associated WH and ZH Production at Hadron Colliders},
    [arXiv:hep-ph/0402003].

\bibitem{14}
    A. Stange, W. Marciano, and S. Willenbrock, Phys. Rev. {\bf D50}
    (1994) 4491, [arXiv:hep-ph/9404247].

\bibitem{15}
    M. L. Ciccolini, S. Dittmaier and M. Kr\"amer, {\it Electroweak Radiative Corrections
    to Associated WH and ZH Production at Hadron Colliders, Phys. Rev.bf D} {\bf 68}
    (2003) 073003, [arXiv:hep-ph/0306234].

\bibitem{tag1a}
    Stefan Soldner-Rembold, {\it Standard Model Higgs Searches}, MAN/HEP/2008/5,
    [arXiv:0803.1451v2].

\bibitem{tag1b}
    Stefan Soldner-Rembold, {\it Standard Model Higgs Searches and Perspectives
    at the Tevatron}, [arXiv:hep-ex/0610014].

\bibitem{tag2}
    D0 Collaboration: V. M. Abazov, et al., {\it A combined search for the
    standard model Higgs boson at $\sqrt{s}=1.96~TeV$, Phys. Lett. B} {\bf 663} (2008) 26,
    [arXiv:hep-ph/0712.0598].

\bibitem{tag3}
    J. M. Butterworth, A. R. Davison, M. Rubin, G. P. Salam, {\it Jet substructure
    as a new Higgs search channel at the LHC, Phys. Rev. Lett.} {\bf 100} (2008)
    242001, [arXiv:0802.2470].

\bibitem{QCDc1}
    T. Han and S. Willenbrock, {\it QCD correction to the $pp \to WH$ and $ZH$ total
    cross sections, Phys. Lett. B} {\bf 273} (1991) 167.

\bibitem{QCDc2}
    J. Ohnemus and W. J. Stirling, {\it Order-$\alpha_s$ corrections to the
    differential cross section for the WH intermediate-mass Higgs-boson signal,
    Phys. Rev. D} {\bf 47} (1993) 2722.

\bibitem{QCDc3}
    H. Baer, B. Bailey and J. F. Owens, {\it $O(\alpha_s)$ Monte Carlo approach to
    W+ Higgs-boson associated production at hadron, Phys. Rev. D} {\bf 47} (1993) 2730.

\bibitem{QCDc4}
    S. Mrenna and C.-P. Yuan, {\it Effects of QCD Resummation on $W^+h$ and $t\bar b$
    Production at the Tevatron, Phys. Lett. B} {\bf 416} (1998) 200, [arXiv:hep-ph/9703224].

\bibitem{QCDc5}
    M. Spira, {\it QCD Effects in Higgs Physics, Fortsch. Phys.} {\bf 46} (1998) 203,
    [arXiv:hep-ph/9705337].

\bibitem{nnlo1}
    O. Brein, A. Djouadi, R. Harlander, {\it NNLO QCD corrections to the Higgs-strahlung
    processes at hadron colliders, Phys. Lett. B} {\bf 579}
    (2004) 149, [arXiv:hep-ph/0307206].

\bibitem{me}
    J. J. Su, W. G. Ma, R. Y. Zhang and L. Guo, {\it Next-to-leading order QCD
    predictions for the hadronic $WH$+jet production, Phys. Rev. D} {\bf 81} (2010)
    114037, [arXiv:1006.0279].

\bibitem{fey}
    T. Hahn, {\it Generating Feynman Diagrams and Amplitudes with FeynArts 3,
    Comput. Phys. Commun.}, {\bf 140} (2001) 418.

\bibitem{form}
    T. Hahn, M. Perez-Victoria, {\it Automatized One-Loop Calculations in 4 and
    D dimensions, Comput. Phys. Commun.}, {\bf 118} (1999) 153.

\bibitem{Passarino}
    G. Passarino and M. Veltman, {\it One-loop corrections for $e^+e^-$ annihilation
    into $\mu^+\mu^-$ in the Weinberg model, Nucl. Phys. B} {\bf 160} (1979) 151.

\bibitem{van}
    G. J. van Oldenborgh and J. A. M. Vermaseren, {\it New Algorithms for One-loop
    Integrals, Z. Phys. C} {\bf 46} (1990)
    425.

\bibitem{R.K}
    R. K. Ellis and G. Zanderighi, {\it Scalar one-loop integrals for QCD,
    JHEP} {\bf 0802} (2008) 002, [arXiv:0712.1851].

\bibitem{Dittmaier}
    S. Dittmaier, {\it Separation of soft and collinear singularities from
    one-loop N-point integrals, Nucl. Phys. B} {\bf 675} (2003) 447, [arXiv:hep-ph/0308246].

\bibitem{tcp}
    B. W. Harris and J. F. Owens, {\it The two cutoff phase space slicing method,
    Phys. Rev. D} {\bf 65} (2002) 094032, [arXiv:hep-ph/0102128].

\bibitem{cteq2}
    D. Stump, J. Huston, J. Pumplin, W. K. Tung, H. L. Lai, S.
    Kuhlmann and J. F. Owens, {\it Inclusive jet production, parton
    distributions, and the search for new physics, JHEP} {\bf 0310} (2003) 046.

\bibitem{hepdata}
    K. Nakamura, et al., (Particle Data Group), {\it Review of Particle Physics,
    J. Phys.G} {\bf 37} (2010) 075021.

\bibitem{jet}
    S. D. Ellis and D. E. Soper, {\it Successive Combination Jet Algorithm For Hadron
    Collisions, Phys. ReV. D} {\bf 48} (1993) 3160, [arXiv:hep-ph/9305266]

\bibitem{Dixon}
    L. J. Dixon, Z. Kunszt, and A. Signer, {\it Vector Boson Pair Production
    in Hadronic Collisions at ${\cal O}(\alpha_s)$: Lepton Correlations
    and Anomalous Couplings,  Phys. Rev. D60}, 114037(1999), [arXiv:hep-ph/9907305].

\bibitem{Uwer1}
    S. Dittmaier, S. Kallweit, P. Uwer, {\it CompHEP 4.4 - Automatic Computations
    from Lagrangians to Events, Phys. Rev. Lett.} {\bf 100} (2008) 062003.

\bibitem{Uwer2}
    F. Campanario, C. Englert, M. Spannowsky, D. Zeppenfeld, {\it NLO-QCD
    corrections to $W\gamma j$ production, Euro. Phys. Lett.} {\bf 88} (2009) 11001.

\bibitem{Uwer3}
    T. Binoth, T. Gleisberg, S. Karg, N. Kauer, G sanguinetti, {\it NLO QCD
    corrections to $ZZ+jet$ production at hadron colliders, Phys. Lett. B} {\bf 683}(2010) 154.

\bibitem{44}
     F. Campanario, C. Englert, S. Kallweit, M. Spannowsky and D. Zeppenfeld, {\it NLO QCD
corrections to WZ+jet production with leptonic decays, JHEP}{\bf 07} (2010) 076,
[arXiv:1006.0390].


\bibitem{CompHEP}
    E. Boos, V. Bunichev, {et al.,} (the CompHEP collaboration), Nucl.
    Instrum. Meth. {\bf A534} (2004) 250-259, [arXiv:hep-ph/0403113].


\bibitem{Berger}
    C. F. Bergera, Z. Bernb, L. J. Dixon, {et al.,} {\it Next-to-Leading Order QCD Predictions for W +3-Jet
    Distributions at Hadron Colliders, Phys. Rev. D } {\bf 80}(2009) 074036.

\end{thebibliography}
\end{document}